\documentclass[conference]{IEEEtran}
\IEEEoverridecommandlockouts
\usepackage{cite}
\usepackage{amsmath,amssymb,amsfonts}
\usepackage{algorithmic}
\usepackage{graphicx}
\usepackage{textcomp}
\usepackage{xcolor}
\usepackage{comment}
\usepackage{caption}
\usepackage{subcaption}
\usepackage{tikz}
\usetikzlibrary{chains,shapes.multipart}
\usetikzlibrary{shapes}
\usetikzlibrary{automata,positioning}
\usetikzlibrary{calc}
\usepackage{tikzscale}
\usepackage{makecell}
\usepackage{wasysym}
\usepackage{enumitem}
\usepackage{soul}
\usepackage{url}
\newcommand{\lecturename}{\textit{Hybrid Quantum-Classical Systems}}
\def\BibTeX{{\rm B\kern-.05em{\sc i\kern-.025em b}\kern-.08em
    T\kern-.1667em\lower.7ex\hbox{E}\kern-.125emX}}
\begin{document}
\title{Training Computer Scientists for the Challenges of Hybrid Quantum-Classical Computing}

\author{\IEEEauthorblockN{Vincenzo De Maio\IEEEauthorrefmark{1}, Meerzhan Kanatbekova\IEEEauthorrefmark{1}, Felix Zilk\IEEEauthorrefmark{1}, Nicolai Friis\IEEEauthorrefmark{2}, Tobias Guggemos\IEEEauthorrefmark{3}\IEEEauthorrefmark{4} and Ivona Brandic\IEEEauthorrefmark{1}}
\IEEEauthorblockA{\IEEEauthorrefmark{1}HPC Laboratory, Institute of Information Systems Engineering, TU Wien, Favoritenstrasse 9-11, 1040 Vienna, Austria}
\IEEEauthorblockA{\IEEEauthorrefmark{2}Atominstitut, TU Wien, Stadionallee 2, 1020 Vienna, Austria}
\IEEEauthorblockA{\IEEEauthorrefmark{3}Faculty of Physics, Vienna Center for Quantum Science and Technology (VCQ), University of Vienna, Vienna, Austria}
\IEEEauthorblockA{\IEEEauthorrefmark{4}Christian Doppler Laboratory for Photonic Quantum Computer, Faculty of Physics, University of Vienna, Vienna, Austria}}

\maketitle 

\begin{abstract}
As we enter the post-Moore era, we experience the rise of various non-von-Neumann-architectures to address the increasing computational demand for modern applications, with quantum computing being among the most prominent and promising technologies. However, this development creates a gap in current computer science curricula since most quantum computing lectures are strongly physics-oriented and have little intersection with the remaining curriculum of computer science. This fact makes designing an appealing course very difficult, in particular for non-physicists. Furthermore, in the academic community, there is consensus that quantum computers are going to be used only for specific computational tasks (e.g., in computational science), where hybrid systems - combined classical and quantum computers - facilitate the execution of an application on both quantum and classical computing resources. A hybrid system thus executes only certain suitable parts of an application on the quantum machine, while other parts are executed on the classical components of the system. To fully exploit the capabilities of hybrid systems and to meet future requirements in this emerging field, we need to prepare a new generation of computer scientists with skills in both distributed computing and quantum computing. To bridge this existing gap in standard computer science curricula, we designed a new lecture and exercise series on \lecturename, where students learn how to decompose applications and implement computational tasks on a hybrid quantum-classical computational continuum. While learning the inherent concepts underlying quantum systems, students are obligated to apply techniques and methods they are already familiar with, making the entrance to the field of quantum computing comprehensive yet appealing and accessible to students of computer science.

\end{abstract}

\begin{IEEEkeywords}
computer science education, post-Moore era, high-performance computing, quantum computing, hybrid systems
\end{IEEEkeywords}

\section{Introduction}

The advent of the post-Moore era confronts scientists of different domains (i.e., scientific computing, high-performance computing (HPC), and machine learning) with the difficulty of scaling their applications beyond the limits of currently available hardware. The scientific community seems to agree that the most promising way to address the challenges of the post-Moore era is the integration of non-von-Neumann-architectures into existing computing infrastructures, where they are envisioned to work as co-processors with classical host processors. Examples of non-von-Neumann-architectures include bio-inspired computing (i.e., neuromorphic computing) as well as quantum computing (QC). As co-processors, these specialized devices receive specific computational tasks with the objective of accelerating the overall computational performance~\cite{HPC_quantum_Humble_2017, bertels2021quantum, Humble_quantum_HPC_2021, Ruefenacht2022BringingQA}.

As a result of a number of recent scientific advances in the field of quantum computing~\cite{Arute2019, Zhong2020, Pogorelov2021, Madsen2022, Castelvecchi2023, Bluvstein2023}, the HPC community is focusing its research efforts on the integration of QC technologies to augment existing HPC infrastructures~\cite{Ruefenacht2022BringingQA, munichquantumvalleyNewsMunich}. Quantum computers promise significant speed-ups for certain computationally intensive problems due to their natural capability to process quantum information. This means that they can be used to solve certain tasks more efficiently than classical computers~\cite{Nielsen2010}. Among the expected fields of application are AI and machine learning~\cite{Cerezo2022}, molecular simulations for material science and drug discovery, and the optimization of supply chains and financial portfolios~\cite{Bayerstadler2021}. This potential acceleration of (scientific) applications in multiple areas makes QC a hopeful candidate to deliver the growing demand for computational performance in the post-Moore era.

However, contrary to typical accelerators already used in HPC (i.e., GPUs and TPUs), QC still requires specific training to fully exploit its potential. First of all, it requires an understanding of quantum mechanical principles that are at the core of the quantum computational paradigm (i.e., superposition states, quantum operations, entanglement, and measurement). Algorithms must therefore be (re)invented, implemented, and tested in the quantum domain~\cite{bertels2021quantum}. Considering that classical data also needs to be encoded into a quantum computer and that inefficient data encoding could potentially cancel out quantum speed-ups, data encoding patterns require particular attention, too~\cite{DataEncodingPatterns2022}. In addition, professionals working in the field of quantum-accelerated computing should possess the ability to evaluate when executing tasks on a quantum computer might improve application performance and which HPC workflows and accelerator architectures are the most efficient~\cite{HPC_quantum_Humble_2017, Humble_quantum_HPC_2021, Ruefenacht2022BringingQA}. It is therefore of paramount importance to train a new generation of computer scientists who are capable of meeting the requirements of modern development tasks for hybrid systems and thus being prepared for the challenges of the post-Moore era.


In this paper, we describe the design of our lecture and exercise series on \lecturename. Our course is an attempt to fill a gap in standard computer science curricula. On the one hand, we want to give students a basic background in quantum computing. On the other hand, we want to provide a unique computer science-oriented perspective that allows them to not only develop quantum algorithms and run them on a quantum computer as an isolated standalone machine, but rather adopt an information systems and architecture perspective. This involves different accelerator architectures, HPC workflows, and the software stack for quantum accelerated computing, and allows students to identify potential performance improvements with quantum computers in order to maximize their use as supercomputing accelerators. Our course consists of a combination of frontal lectures and practical exercises with real or simulated quantum hardware, which is made available to us through a cloud frontend for quantum hardware.

The paper is organized as follows: First, we outline the design of the course, including intended learning outcomes, accompanying literature, the course syllabus, and a description of the lecturers' profiles in Section~\ref{sec:design}. Afterwards, we briefly explain the main topics that will be presented in the course on \lecturename~in Section~\ref{sec:concepts}. In Section~\ref{sec:evaluation}, we clarify how the students are evaluated. The typical infrastructure in which they are expected to work is described in Section~\ref{sec:infrastructure}. In Section~\ref{sec:eval} we evaluate our course, while in Section~\ref{sec:lessonslearned} we discuss our experience in teaching the course within the computer science curriculum of TU Wien. In Section~\ref{sec:related} we highlight related work, and we conclude our paper in Section~\ref{sec:conclusion}.

\section{Course Design}
\label{sec:design}

\subsection{Learning Outcomes (LO)}
\label{sec:learningoutcomes}
The course is designed for master's students in computer science and related subjects at TU Wien and is intended to build upon the students' already acquired knowledge and skills in standard topics of computer science (such as computer architecture, theoretical computer science, and software engineering) and expand these to include concepts and methods for the use of hybrid quantum-classical systems, i.e., the integration of quantum accelerators into HPC systems. We expect mathematics as well as computer science knowledge at the bachelor's level in computer science, but do not assume any prior knowledge of (quantum) physics.

After successfully completing the course, students are able to, in particular:

\begin{itemize}
    \item LO1: Explain how the physics of quantum computation is different from classical computational models;
    \item LO2: Describe the theoretical performance improvements that (hybrid) quantum algorithms offer compared to classical algorithms;
    \item LO3: Analyze the life cycle of hybrid applications and decompose their execution on a hybrid quantum-classical
    computational continuum;
    \item LO4: Develop their own (hybrid) quantum algorithms and implement them using (real or simulated) quantum computers using quantum toolkits such as Qiskit \cite{ibmQuantumComputing}.
\end{itemize}

The learning outcomes (LO1 - LO4) are defined according to Bloom's taxonomy~\cite{andersonkrathwohl} and a student-centered outcome-based methodology~\cite{learningoutcomesguide}. They also specify the expected outcomes and requirements for students of this course.

\subsection{Proposed Textbook}
When choosing the accompanying literature, we selected the following book by N. S. Yanofsky and M. A. Mannucci, \textit{Quantum Computing for Computer Scientists} \cite{yanofsky_mannucci_2008}, from an existing, rich portfolio of specialist literature. This book serves as a bridging element between traditional computer science and introductory quantum computing. Its presentation is particularly accessible for computer science students since the text draws on concepts and techniques that are already familiar to the students (e.g., computer architecture, algorithms, theoretical computer science, etc.) and comprehensively builds the core ideas of quantum computation upon them.

Also, we integrate concepts of the textbook with online resources, such as research papers and online documentation (i.e., Qiskit official documentation, IBM Quantum Learning~\cite{ibmQuantumLearning}).

\subsection{Syllabus}
The curriculum begins with an introductory lecture intended to provide an overview of non-von-Neumann-architectures for the post-Moore era, with a specific focus on quantum computing. Then, two subsequent lectures teach the mathematical and physical fundamentals of quantum information, which form the basis for subsequent lectures on quantum circuits and quantum algorithms. Building on this, we focus on selected proof-of-concept as well as hybrid quantum algorithms and their execution on a hybrid quantum-classical computational continuum. We offer separate lectures on the subject areas of variational quantum algorithms (VQA)~\cite{Peruzzo2014}, the quantum approximate optimization algorithm (QAOA)~\cite{PhysRevX.10.021067}, and quantum machine learning (QML)~\cite{DunjkoBriegel2018, Cerezo2022}. Two guest contributions on specialized advanced topics, namely the realization of QC protocols with photonic qubits~\cite{Couteau2023} and quantum error correction techniques~\cite{Devitt2013}, conclude the course.

The teaching goals for each of our lectures are as follows:\\

\subsubsection{Introduction to \lecturename}
In this first class, we outline the syllabus for the course, explain the class setup as a combination of lectures and exercises, and present its intended learning outcomes (see \ref{sec:learningoutcomes}) and evaluation. We also show examples of state-of-the-art realizations of quantum computers with different hardware and explain their relevance in the context of HPC~\cite{HPC_quantum_Humble_2017, bertels2021quantum, Humble_quantum_HPC_2021, Ruefenacht2022BringingQA}.\\

\subsubsection{Fundamentals of Maths for Quantum Computing}
We use this lecture to activate students' prior knowledge of concepts relevant for an understanding of the mathematics of QC. We cover complex numbers, vector spaces, projections, matrices, tensor products, and probabilities~\cite{yanofsky_mannucci_2008, Nielsen2010}.\\


\subsubsection{Basics of Quantum Information}
This lecture introduces the students to the physics of quantum information, acquainting them with quantum superposition states of single and multiple qubits, quantum gate operations, entanglement and correlations, and the role of measurement in quantum theory~\cite{Audretsch2007, Nielsen2010}.\\

\subsubsection{Circuits and Proof-of-Concept Quantum Algorithms}
In this hands-on lecture, we introduce students to quantum computer programming with the IBM Qiskit framework and show them typical execution models for quantum programs as well as how to design small circuits. We discuss the concepts of sampler and estimator interfaces and present the Deutsch-Jozsa algorithm as a proof-of-concept of a quantum algorithm.\\

\subsubsection{Fundamental Quantum Algorithms}
As a follow-up to the previous lecture, we continue to examine more complex quantum algorithms, with a special focus on Simon's periodicity algorithm and Grover's algorithm~\cite{Nielsen2010}. We also demonstrate their practical implementations and execute the algorithms on Qiskit backends.\\

\subsubsection{Variational Quantum Algorithms}
In this lecture, we study the variational principle and the concept of annealing, which constitute the foundations of VQAs~\cite{Cerezo2021}. We discuss the hybrid execution model of VQAs and include the variational quantum eigensolver (VQE) as an example.\\

\subsubsection{Quantum Optimization, QAOA and MaxCut}
We continue the analysis of VQAs by focusing on their applications to combinatorial optimization problems. We focus primarily on the MaxCut problem and show how various instances of MaxCut can be solved using VQAs, especially with QAOA.\\

\subsubsection{Introduction to Quantum Machine Learning}
We review typical applications of QC to machine learning problems, with special attention to clustering problems. We demonstrate how typical clustering algorithms, such as K-means, K-medians and SVM, may be augmented by means of QC.\\

\subsubsection{Quantum Neural Networks}
We show how to train neural networks on quantum computers, exploiting quantum effects to enhance the quality of the training and advance the exploration of the target solution space. We also introduce quantum reservoir computing and variational quantum classifiers and regressors.\\

\subsubsection{Photonic Quantum Computing}
In this guest lecture, we present linear-optics setups as one promising hardware platform for quantum computing and show how qubit states are interpreted as quantum states of light.
We implement a quantum algorithm from the circuit model with optical elements like beam splitters, phase shifters, or wave plates and evaluate why such a direct translation is inefficient.
Next, we introduce measurement-based QC (MBQC)~\cite{Briegel2009} or One-Way QC as a concept that is considered valuable for photonic quantum computing. We describe how this approach is equivalent to gate-based quantum computing, and how one may efficiently translate between the two models using the language of ZX-Calculus~\cite{Coecke_2011,coecke_kissinger_2017}.\\

\subsubsection{Quantum Error Mitigation}
After briefly motivating the need for as well as the challenges to quantum error correction, we review the basics of error detection and error correction within the stabilizer formalism. We introduce the students to relevant concepts such as logical operators, stabilizers, parity measurements, 
and code distance, using the three-qubit repetition code as a starting example before moving on to topological surface codes~\cite{Kitaev2003}. Possible implementation with trapped ions will be briefly discussed.\\

\subsection{Lecturers}
The lecturers of this first iteration of the course from the winter semester 2023/24 are:

\begin{description}
    \item[Vincenzo De Maio] has a PhD in computer science with a background in cloud computing and resource allocation. He performs research at the interface between high-performance computing and quantum computing, including hybrid quantum algorithms and the HPC-QC software stack. He is responsible for coordinating the lecture and preparing the assignments in collaboration with the other lecturers.\\ 
    \item[Meerzhan Kanatbekova] holds a master's degree in computer science and a bachelor's degree in applied mathematics. She is responsible for the lectures on the mathematical prerequisites and the lecture on combinatorial optimization problems.\\ 
    \item[Felix Zilk] holds a master's degree in physics with a background in experimental quantum optics and quantum information science. During his master's, he also investigated measurement-based quantum computing protocols and their software frameworks for experimental realizations with photonic qubits.\\ 
    \item[Tobias Guggemos] holds a PhD in computer science and has been teaching quantum computing to computer science students for several years. He is researching quantum algorithms using photons, including novel experimental architectures for measurement-based quantum computing, quantum machine learning on integrated photonic chips, secure quantum cryptography and ''Beyond QKD`` protocols with quantum dots and nonlinear crystals.\\  
    \item[Nicolai Friis] holds a PhD in mathematical physics and has completed a habilitation (venia docendi) in theoretical physics, with more than a decade of teaching and research experience in quantum information, quantum computing and other theoretical-physics topics. He will introduce the students to quantum error correction. \\
    \item[Ivona Brandic] is a full professor in computer science with experience in HPC systems, scientific computing and workflows. She has more than  a decade of experience in teaching distributed systems and other HPC-related topics. She will provide lectures on future developments in hybrid quantum-classical systems and post-Moore computing. \\
\end{description}

The diverse backgrounds of the teachers reflect the high level of interdisciplinary work required in this class. We hold the view that having access to a wide range of specialist knowledge is an advantage, but we do not consider such a variety of expert instructors to be necessary to carry out the course as described.

\section{Main Topics in the Lecture}
\label{sec:concepts}

Here we group the main topics that are presented in the lecture. The lecture materials (i.e., slides, assignments, etc.) are accessible through a password-protected cloud folder on the TU Wien Owncloud, at the url \url{https://vindem.github.io/teaching/hcqs/}\footnote{We provide the password upon request via email to hcqs-lecture@ec.tuwien.ac.at.} 
.
 
\subsection{Fundamentals of Quantum Information}
As fundamentals of quantum information for computer scientists, the lecture covers quantum superposition states of single and multiple qubits, quantum gate operations, entanglement and correlations, and the description of measurements in the context of quantum mechanics.

The lecture places great importance on understanding the concept of quantum superposition. Here, it is extremely important that students learn the physical difference between the quantum superposition state of a qubit and the stochastic state of a classical probabilistic bit. In many common introductions, the superposition state of a qubit (with respect to the computational basis) is explained as "0 and 1 at the same time", or similar, which we consider to be inadequate and misleading.

The character of quantum superposition can be highlighted by measuring the state of a qubit in different bases, i.e., corresponding eigenbasis~\cite{dür2013learn}. Thereby, we point out the destructive role and the intrinsic randomness of the measurement process in quantum mechanics, by which the state vector of a qubit is randomly projected onto one of the basis vectors.

We use the Bloch sphere~\cite{Nielsen2010, dür2013learn} as a visualization tool for depicting quantum superposition states and the measurement of a qubit's state, and for analyzing quantum gate operations.

We encounter entangled states in the lecture as soon as we consider systems with more than two qubits. Here, we take a look at the correlations of the measurement results upon measuring entangled states and compare them with the correlations of classical (mixed) states~\cite{Audretsch2007}.

\begin{figure}[!ht]
    \centering
    \includegraphics[width=.9\columnwidth]{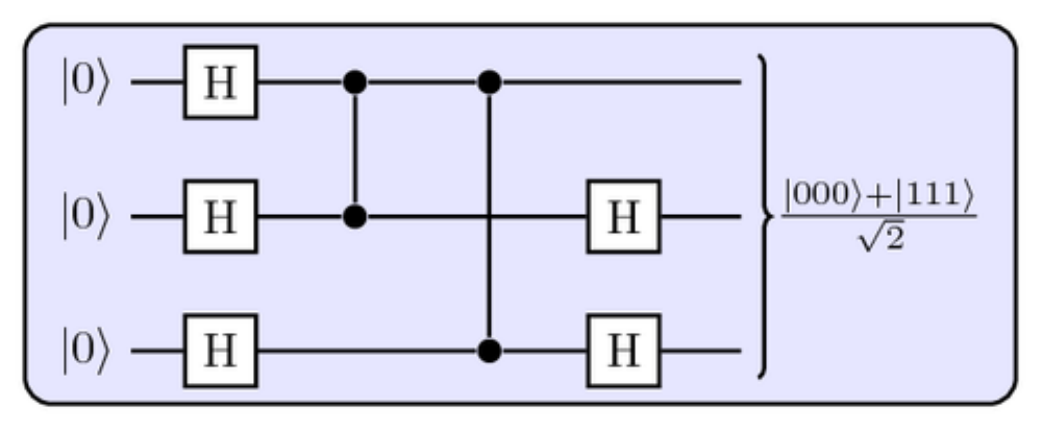}
    \caption{Example of a quantum circuit. Each horizontal line represents a qubit, with the (temporal) order of operations progressing from left to right. Here, all qubits are initialized in the state "$\left|0\right\rangle$", before a sequence of single-qubit Hadamard gates (white boxes, "H"), and two-qubit controlled-$Z$ gates (vertical lines with black circles indicating the respective qubit pairs) is applied. The resulting three-qubit state is the (genuinely multipartite entangled) Greenberger-Horne-Zeilinger (GHZ) state $\bigl(\left|000\right\rangle+\left|111\right\rangle\bigr)/\sqrt{2}$.}
    \label{fig:quantum-circuit}
\end{figure}

\subsection{Programming Models for Quantum Computing}
We describe the two main programming models that students will learn in the class.

\subsubsection{Gate-Based}

In the gate-based model~\cite{Nielsen2010} (also known as circuit model), a quantum algorithm is expressed as a circuit. A circuit, or quantum circuit, consists of one or multiple qubits initialized to a specific state and a layer of operations, so-called quantum gates. The quantum gates manipulate the state of one or multiple qubits and thereby carry out the quantum logical operation. An example of a quantum circuit is shown in Figure~\ref{fig:quantum-circuit}. In our lecture, we focus mainly on the gate-based model, since it is the standard approach taken towards QC, has direct correspondence with classical electrical circuits, which computer scientists are already familiar with from their exams in computer architecture.

\subsubsection{Measurement-Based}
The measurement-based-, or one-way model~\cite{Briegel2009} describes the manipulation of quantum states as measurements on entangled states and subsequent corrections on the remaining qubits conditioned on the outcomes of these measurements. In this regard, quantum algorithms are implemented as a series of adaptive single-qubit measurements and correction gates on an entangled state (called \textit{resource state}).
This model is not part of the exam material; however, students should know that there are many distinct ways of processing quantum information. The measurement-based model is particularly interesting due to its prominence in various sub-areas of quantum information science and its practical advantages for certain physical setups.

\subsection{Execution Models for Quantum Computation}

\begin{figure*}[!ht]
    \centering
    \begin{subfigure}[b]{0.45\textwidth}
    \centering
    \includegraphics[width=\textwidth]{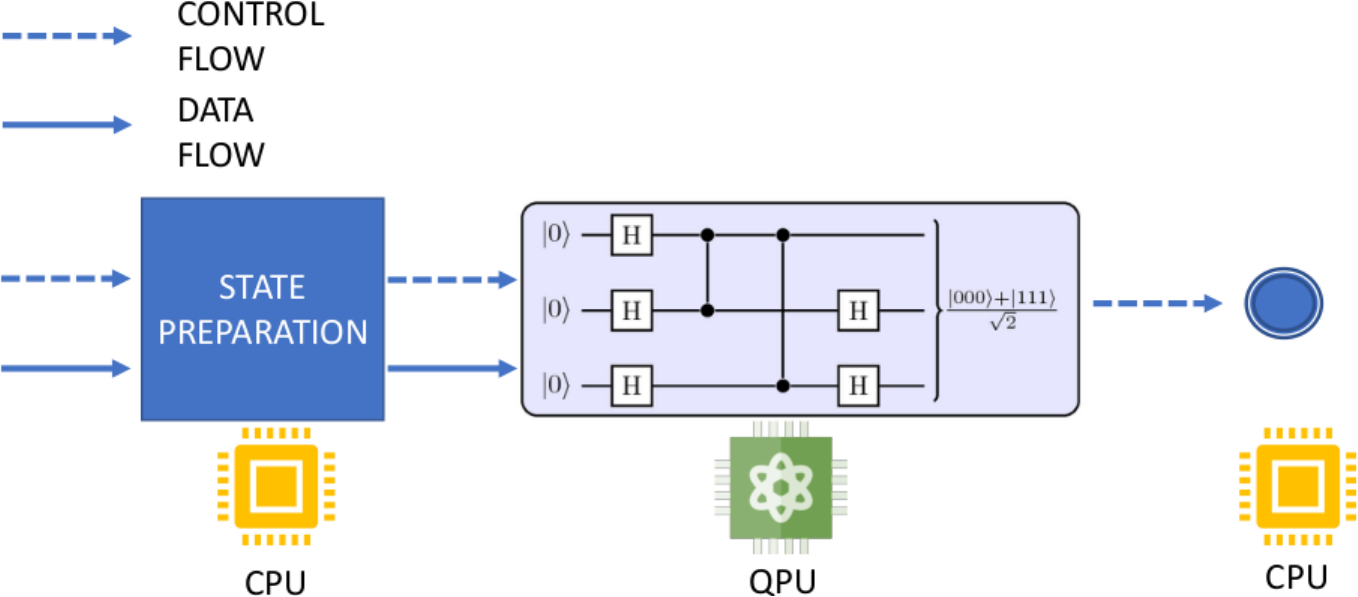}
    \caption{Single Execution Model}
    \label{fig:single-execution}
    \end{subfigure}
    \begin{subfigure}[b]{0.45\textwidth}
    \centering
    \includegraphics[width=\textwidth]{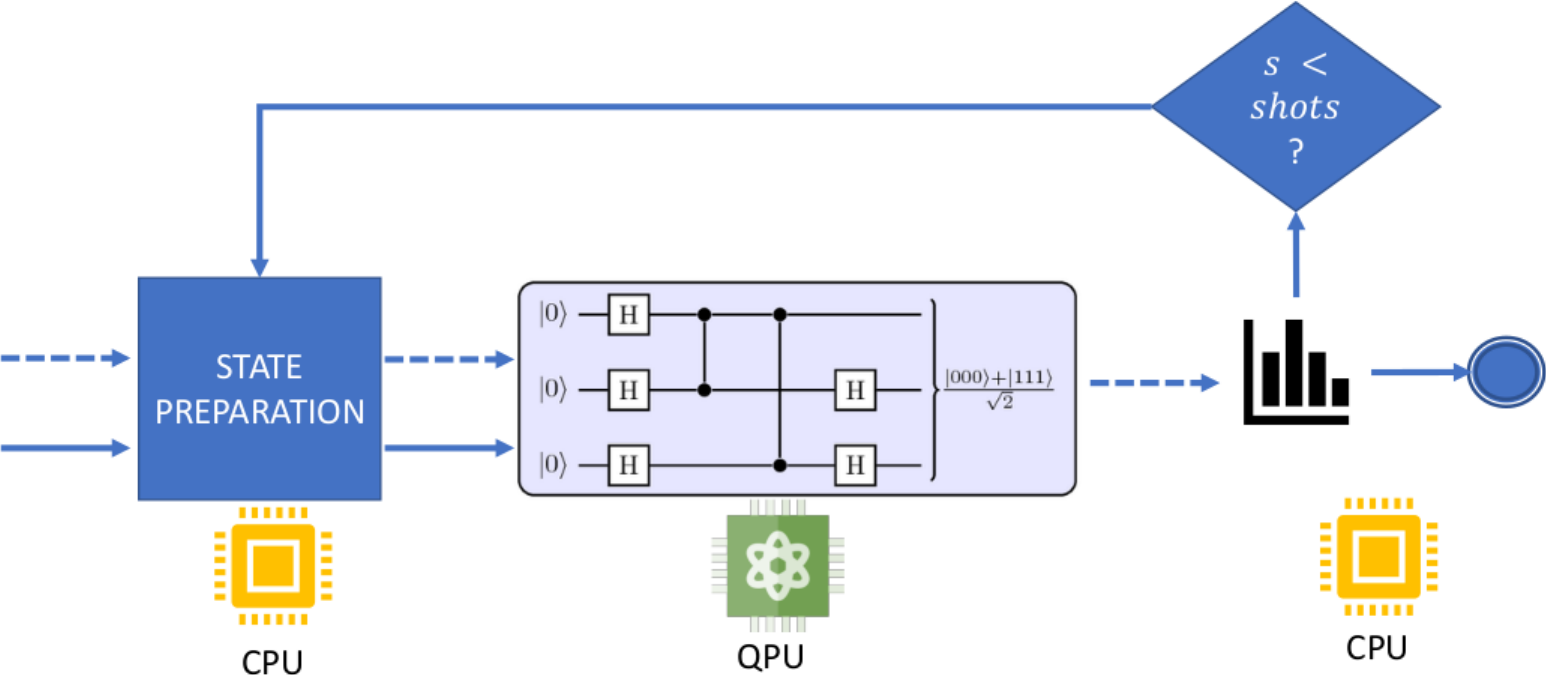}
    \caption{Job Execution Model}
    \label{fig:job-execution}
    \end{subfigure} \\
    \begin{subfigure}[b]{0.45\textwidth}
    \centering
    \includegraphics[width=\textwidth]{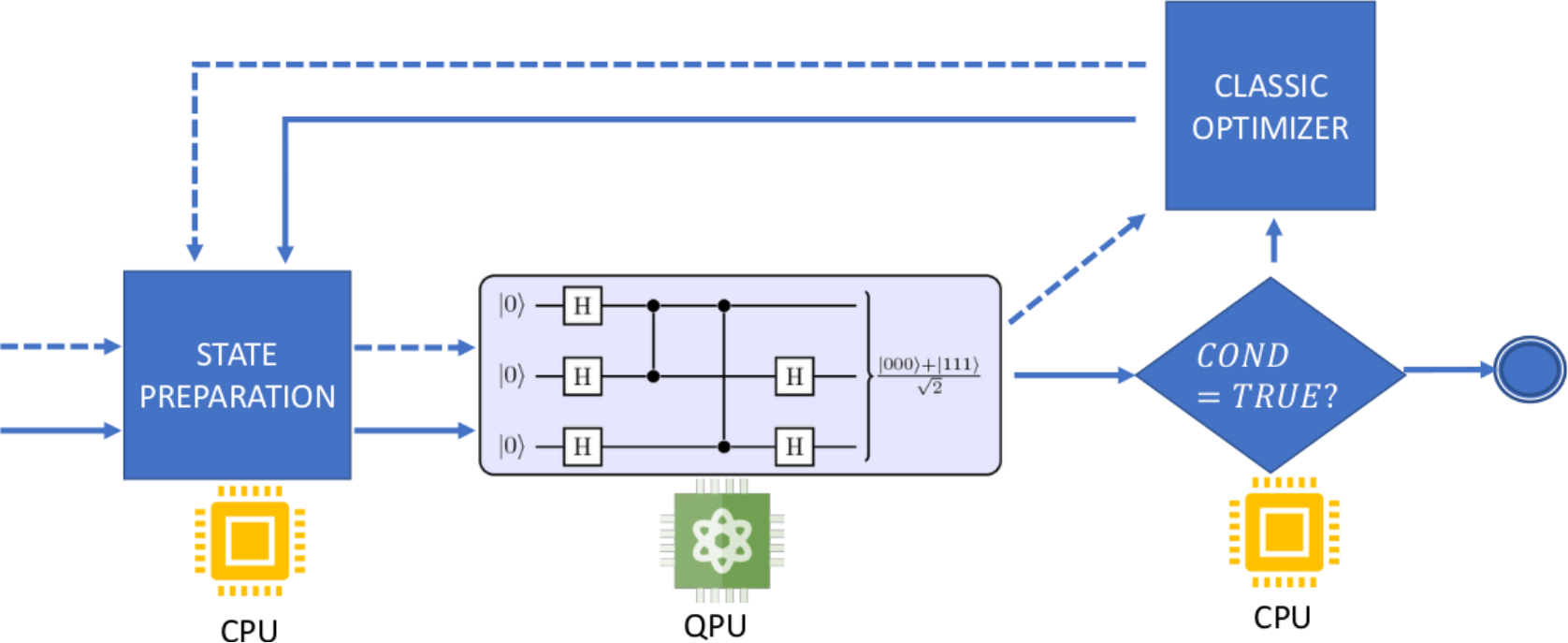}
    \caption{Hybrid Execution Model}
    \label{fig:hybrid-execution}
    \end{subfigure}

    \caption{After completing the course, students should be able to describe and implement the depicted execution models for quantum computation. The (a) Single Execution Model performs one single run of a quantum circuit, whereas the (b) Job Execution Model carries out the same circuit repeatedly to obtain statistics of different outcomes from which a solution is interpreted. In both models, the classical computers involved are used to interface with and control the quantum system. The (c) Hybrid Execution Model further relies on classical computers to execute computational parts of an algorithm, such as optimization subroutines in VQAs.}
    \label{fig:execution-models}
\end{figure*}

At the end of the class, students should know how to develop and implement hybrid quantum-classical applications using different quantum task execution models. These execution models are depicted in Figure~\ref{fig:execution-models} and defined as follows:

\begin{itemize}
    \item \textbf{Single Execution (see Figure~\ref{fig:single-execution})} models the single execution of a quantum circuit, which is terminated by the measurement of its outcome;
    \item \textbf{Job Execution (Figure~\ref{fig:job-execution})} represents the repeated execution of the same quantum circuit. Execution is repeated for a pre-defined number of times (\textit{shots}). The final result of this execution can be either the expectation value or the quasi-probability distribution of the outcomes;
    \item \textbf{Hybrid Execution (Figure~\ref{fig:hybrid-execution})} refers to the execution model of quantum computations that split a computational workload between quantum and classical computing resources, as, for example, in VQAs, where results of classical computations are used to drive the execution on the quantum device~\cite{Cerezo2021}.
\end{itemize}

The classical computers involved in both the Single and Job Execution Model are employed to interface with the quantum machine and operate the quantum processor's control systems; however, they do not carry out any computational parts of the executed algorithm. In contrast, the Hybrid Execution Model integrates classical computers as parts of the overall computation.

\subsection{Variational Quantum Algorithms}

\begin{figure}[h]
    \centering
    \tikzstyle{startstop} = [rectangle, minimum width=0.3cm, minimum height=0.3cm,text centered, draw=black, fill=red!30]
    \tikzstyle{process} = [rectangle, minimum width=1cm, minimum height=0.3cm, text centered, draw=black, fill=blue!30]
    \tikzstyle{decision} = [diamond, minimum width=0.3cm, minimum height=0.3cm, text centered, draw=black, fill=white!30]
    \tikzstyle{arrow} = [thick,->,>=stealth]
    \begin{tikzpicture}
    \node (start) [startstop] {Start};
    \node (initial) [process, right=.2cm of start] {\makecell[c]{Initial\\$\vec{\Theta}$}};
        \node (quantum) [process, right=.3cm of initial] {\makecell[c]{PQC\\Execution}};
    \node (decision) [decision, right=.3cm of quantum] {\makecell[c]{Found\\$\vec{\Theta^*}?$}};
    \node (c-theta) [process, below=.4cm of decision] {\makecell[c]{Optimize\\$C(\vec{\Theta})$}};
    \node (new-theta) [process, left=.4cm of c-theta] {\makecell[c]{New\\$\vec{\Theta}$}};
    \node (stop) [startstop, right=.85cm of decision] {Stop};
    \draw[arrow] (start) -- (initial);
    \draw[arrow] (initial) -- (quantum);
    \draw[arrow] (quantum) -- (decision);
    \draw[arrow] (decision) -- node[anchor=west]{\texttt{NO}} (c-theta);
    \draw[arrow] (c-theta) -- (new-theta);
    \draw[arrow] (new-theta) -| (quantum);
    \draw[arrow] (decision) -- node[anchor=south]{\texttt{YES}} (stop);
    \end{tikzpicture}
    
    \caption{Variational quantum algorithms.}
    \label{fig:vqas}
\end{figure}
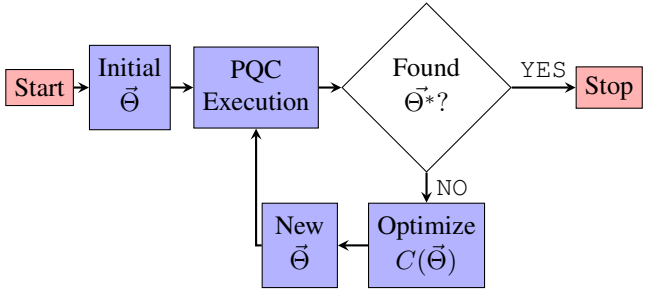

Variational quantum algorithms (VQAs) are a class of algorithms that rely on a hybrid quantum-classical loop, in which a classical optimizer is used to train a parametrized quantum circuit (PQC). The key idea behind VQAs is to minimize a parametrized cost function, $C$, that encodes the solution to a given problem (e.g., the energetic ground state $H_G$ of a molecule with Hamiltonian $H$). In each iteration of the loop, this cost function is evaluated on the quantum computer, followed by classical optimization of the function parameters $C(\vec{\Theta})$~\cite{Cerezo2021}, as depicted in Figure~\ref{fig:vqas}.

VQAs are of high importance to our course since they provide a promising approach towards the realization of quantum advantages. Also, they potentially allow one to reduce the number of qubits with respect to other comparable quantum algorithms and to fully exploit the capabilities of hybrid quantum-classical systems. Moreover, they are used in a wide set of applications that are typical in HPC, such as the simulation of molecular dynamics~\cite{cranganore2022}, the study of material properties~\cite{dallairedemers2018lowdepth}, and matrix multiplication~\cite{Xu_2021}. In this lecture, we demonstrate to the students how to apply VQAs in different applications. For combinatorial optimization, we focus on the QAOA, which is often used for combinatorial optimization problems, and the variational quantum eigensolver (VQE), which is used to compute matrix eigenvalues to find the ground state energy of a Hamiltonian~\cite{Cerezo2021}. Afterwards, we show applications of VQAs for machine learning, with a special focus on quantum neural networks~\cite{Cerezo2022}.

\subsection{Design of Hybrid Applications}

Finally, we show students how to decompose a classical HPC application, modelled as a scientific workflow, into a hybrid quantum-classical application. The abstraction of scientific workflows is well-known in the HPC scientific community since it offers a straightforward way to decompose applications into interdependent tasks. Using a workflow management system such as Pegasus~\cite{Deelman2021}, each job can then be allocated to different nodes within an HPC infrastructure. 

\begin{figure*}[!ht]
    \centering
    \begin{subfigure}{\textwidth}
        \includegraphics[width=\textwidth]{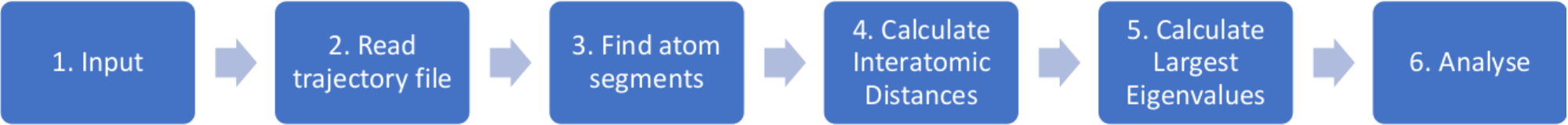}
        \caption{Initial workflow.}
        \label{fig:initial-workflow}
    \end{subfigure}
    \begin{subfigure}{\textwidth}
        \includegraphics[width=\textwidth]{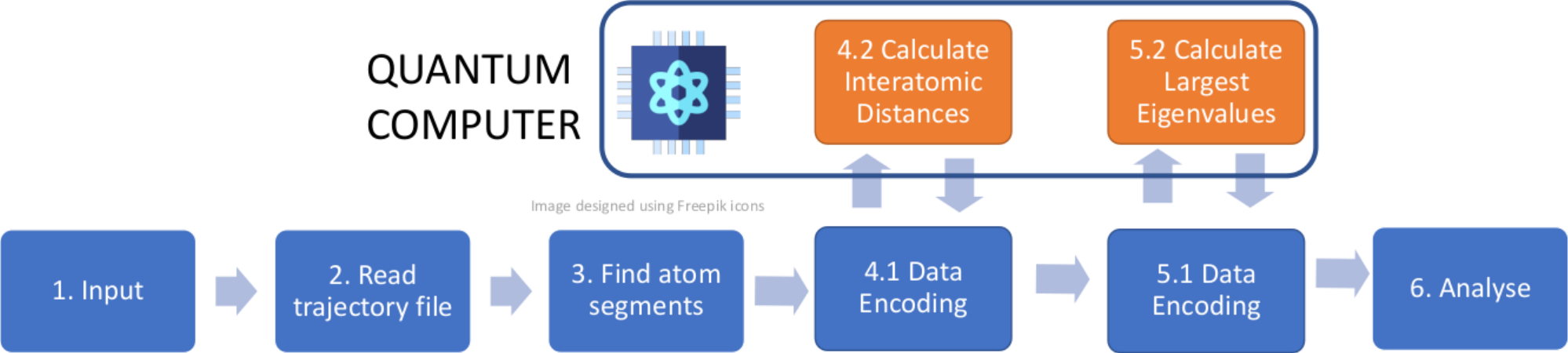}
        \caption{Hybrid quantum-classical workflow.}
        \label{fig:hybrid-workflow}
    \end{subfigure}
    \caption{Scientific workflow decomposition into a hybrid quantum-classical workflow. Here, for example, the computation of interatomic distances and largest eigenvalues in the (a) initial workflow are identified as suitable candidates for implementation on a quantum computer in a (b) hybrid workflow.}
    \label{fig:hybrid-workflow-decomposition}
\end{figure*}

In our course, students begin with a workflow description and identify which tasks are suitable candidates for a quantum implementation before designing a decomposition into a quantum-classical workflow. An example of a workflow decomposition is provided in Figure~\ref{fig:hybrid-workflow-decomposition}. 

\section{Students' Evaluation}
\label{sec:evaluation}
Three homework assignments are designed to deepen the concepts taught. The first two assignments are individual. The last assignment is a group assignment and covers all the concepts required for the development of hybrid quantum-classical applications.

\subsubsection{Assignment 1: Basics of Quantum Computing}


This assignment requires the calculation of tensor products and unitaries, and measurement bases, superposition, and entanglement have to be explained. The assignment is composed of:
\begin{itemize}
    \item Two theoretical exercises about orthonormal bases and unitaries;
    \item Two theoretical exercises about measurement bases and entanglement;
    \item Two programming exercises on superposition and entanglement.
\end{itemize}

The goal of this assignment is to evaluate whether students understand fundamental concepts of quantum information theory and are able to apply them in the design of quantum programs.
\subsubsection{Assignment 2: Quantum-Augmented K-Means and SVM Classification}
The scope of this assignment is to assess the students' understanding of the gate-based model (see Figure~\ref{fig:quantum-circuit}) and the different execution models (see Figure~\ref{fig:execution-models}). The main goal is to take an algorithm and identify which parts could be improved if executed on quantum devices, implement these parts, and integrate the result into the application execution. We selected K-means as a target application, as it is discussed in the lecture about QML. We also let students perform classification using quantum SVM to gain experience in applying QML to real datasets. The assignment is composed of three exercises:
\begin{itemize}
    \item \textbf{Exercise 1: Design of classical K-Means algorithm.} The goal of this exercise is to design the baseline implementation of classical K-means that will be used as a comparison for quantum-augmented K-means.
    \item \textbf{Exercise 2: Quantum-Augmented K-Means.}
    The goal of this exercise is to implement a version of K-Means that is augmented with quantum tasks. Students are required to (1) identify which tasks of K-means could benefit from execution on quantum hardware, (2) identify a quantum algorithm for each identified task, and (3) implement at least one version as hybrid-quantum algorithms. Students may choose not to implement any quantum algorithm, as long as they provide justification for their decision.
    \item \textbf{Exercise 3: Quantum SVM.} In this exercise, students should perform a classification task using a well-known classification dataset from the \texttt{scikit} package. The goal is to gather experience with the encoding of classical data into quantum data and perform data analysis using quantum SVM classification.
\end{itemize}
\subsubsection{Assignment 3: Bin-Packing Optimization}
This group assignment focuses on combinatorial optimization on quantum hardware using the QAOA or VQE algorithm. The target problem is bin-packing, which has different applications to computer science problems such as multiprocessor scheduling and logistics optimization. Groups are composed of maximum of three students. Each group will provide an implementation of a bin-packing problem (BPP) with a quantum computer, considering all the following phases:
\begin{enumerate}
    \item \textbf{Preprocessing}: Parsing each instance of BPP that will be provided by the lecturers as a specific dataset;
    \item \textbf{QUBO transformation}: Transforming each instance into Quadratic Unconstrained Binary Optimization (QUBO) form using different penalty methods;
    \item \textbf{Ising Hamiltonian transformation}: Transformation of QUBO into an Ising Hamiltonian;
    \item \textbf{Encoding}: Encoding of an Ising Hamiltonian as quantum data;
    \item \textbf{Solution}: Solving encoded instances using QAOA or VQE; 
    \item \textbf{Analysis}: Analysis of the quantum solution compared to the classical solution of each instance, applying different metrics (i.e., distance from optimum, percentage of times the quantum algorithm achieves the optimum).
\end{enumerate}

\section{Target Infrastructure}
\label{sec:infrastructure}
Since the lecture and exercise series focuses on the design of hybrid quantum-classical systems, we give students access to the HPC infrastructure of our university to perform their exercises. Also, since they are required to run applications on real quantum devices, we need to select a quantum programming framework that supports execution with different quantum backends. We selected IBM Qiskit since it provides sufficient didactic resources, including an online textbook and a series of tutorials to support the students in their learning~\cite{ibmQuantumLearning}.

Regarding access to quantum computers, we provide access to a simulator running in our laboratory, a RasQberry device offered to us by IBM. This device can be used to augment the simulation capabilities offered by the students' personal hardware. Also, we let the students experience execution on real quantum machines through the open plans of providers such as IBM Quantum~\cite{ibmQuantumComputing} and Amazon BraKet~\cite{amazonQuantenCloudComputingServiceAmazon}.





\section{Evaluation}
\label{sec:eval}

We want to point out is that the course was well received by the students: among the $35$ students who enrolled in the class, the final assignment was taken by $19$ students. To this number, we have to add $3$ students who joined the class as auditors and did not take the exam. The total drop-out rate is then about $35$\%, which is in line with the average of TU Wien for elective classes. Among the students, only two completed the class evaluation. The feedback is very positive, but unfortunately we cannot draw reliable conclusions from such as small sample.

\begin{table}[h]
\centering
\begin{tabular}{|c|@{}c@{}|c|c|c|@{}c@{}|}
\hline
Learning  & & Assignment & Assignment & Assignment & Ratio\\
Outcome & & 1 & 2 & 3 & \\ 
\hline
\hline
LO1              &MAX    & $2200$ & \textit{n.a.} & $1900$  & $90.1$\%  \\
                 \cline{2-5} 
                 &SCORED &  $2105$     & \textit{n.a.} &  $1590$&  \\
                 \hline
LO2              &MAX    & \textit{n.a.} & $1425$  &  $1900$    & $88.2$\% \\
                 \cline{2-5}
                 &SCORED   & \textit{n.a.} &  $1340$  &  $1590$ &  \\
                 \hline
LO3              &MAX    & \textit{n.a.} & $855$   &    $1900$ & $85.8$\% \\
                 \cline{2-5}
                 &SCORED   & \textit{n.a.} & $775$   &   $1590$  & \\
                 \hline
LO4              &MAX    &   $770$      & $1330$  &  $1900$  & $88.3$\% \\
                \cline{2-5}
                &SCORED   &   $675$       &  $1270$  & $1590$ & \\
\hline
\end{tabular}
\caption{Percentage of Points per LO per Assignment.}
\label{tab:lo-students}
\end{table}

We also performed an evaluation to quantify (as percentages) to which degree students reached the learning outcomes described in Section~\ref{sec:learningoutcomes}. The results are summarised in Table~\ref{tab:lo-students}. For each assignment we consider the maximum  amount of points available for the whole class per LO and the amount of points that the class scored (collectively) for each LO, while \textit{n.a.} (\textit{not available}) indicates that the corresponding LO has not been evaluated through this assignment. Thus, LO1 has been measured through Assignments 1 and 3, LO2 through Assignments 2 and 3., etc. 

The results show that the LOs have been reached with percentages ranging from $85$\% to $90$\%, which we consider to be evidence that the course has reached its main objectives. Since LO3 seems to have been the most challenging for the students, with only $85$\% achieved, we intend to improve the course by including additional emphasis and exercises targeting LO3 in the future. 

\section{Lessons Learned}
\label{sec:lessonslearned}
Overall, we learned the following lessons:

\begin{itemize}
    \item For this edition of the lecture, we decided to perform an \textbf{evaluation of the students only based on the results of their assignments}. On the one hand, this 
    fits in well with 
    the hands-on approach that we wanted to encourage. 
    On the other hand, it was difficult to cover the topics of all lectures with the assignments, especially concerning the guest lectures. Therefore, in the future editions of the class, we aim to include a written or a oral exam in order to cover the topics presented in the lecture more uniformly.
    
     \item Another lesson learned concerns the difficulty of presenting \textbf{an appropriate amount of quantum-mechanics concepts} required for the understanding of quantum computing whilst keeping the focus of the class on a practical viewpoint. 
     In this first edition of the lecture, we chose to include a limited set of theoretical concepts to facilitate understanding by master students in computer science curriculum, which forced us to leave out concepts such as quantum teleportation and to limit the discussion on computational universality and other important theoretical results. However, considering that the topics introduced in the lecture (see Section~\ref{sec:concepts}) were well received by the students, we will consider to increase the amount of quantum physics concepts presented in the lecture, in order to expand the course program and provide more in-depth insights on the aforementioned notions. To this end, we might include more material from Quantum Information and Computation by Nielsen and Chuang~\cite{Nielsen2010}, which has a more physics-oriented view on these topics. 

     \item Only a small number of students completed the evaluation of the course. Therefore the feedback to improve the class that we have received so far is limited. As the course evaluation is not mandatory at our institution, we aim to further \textbf{motivate the students to participate in the course evaluation} in order to collect more tailored feedback on how to improve this course.
\end{itemize}

\section{Related Work}
\label{sec:related}
The landscape of higher education at the intersection of classical and quantum computing systems receives increasing attention and is therefore expected to grow further in the years ahead. Existing courses are often tailored to distinct audiences and offer introductory quantum mechanics and quantum computing programs to computer scientists~\cite{fagonzalezoQuantumComputer, helsinkiStudies, indianaprogrammingqc, wrfranklinQuantumECSE49646964, isFIPV275Intro, gatech8803O13Quantum} or teach programming skills and computer science concepts to trained physicists~\cite{cvutQuantumProgramming}. These courses are based on hybrid systems in the sense that quantum algorithms are executed on simulators or cloud-based quantum computing services, which require standard classical computers to enable access to and to control the quantum processor.

The course \textit{Quantum Computer Programming}~\cite{stanford269QElements} taught by D. Boneh and W. Zeng at Stanford University in 2019 shares similarities with the curriculum presented in this paper. Their course places an emphasis on the programming of QC systems. Moreover, it explicitly introduces the topic of benchmarking quantum systems, which goes beyond the scope of this first iteration of our course.

In contrast, the Deggendorf Institute of Technology has launched an entire three-semester master's program dedicated to high-performance computing and quantum computing~\cite{thdegHighPerformanceComputing}. This program involves a full spectrum of subjects ranging from fundamental mathematics and (quantum) physics to software and networking technologies, the design and operation of hybrid computer architectures, and the deployment of hybrid applications relevant to quantum-accelerated HPC.

\section{Conclusion and Future Work}
In this paper, we have described our efforts and experiences in designing a master's-level lecture and exercise series that introduces quantum computing in the context of information systems, in particular HPC integration, to students in the fields of computer science and related subjects at TU Wien.

The scientific community predicts a strong impact on the technological ecosystem in the HPC environment in particular, and quantum computers may require new types of infrastructure and designs of information systems architectures.

Our contribution lies in offering a comprehensive yet accessible lecture and exercise series, including individual homework and project work, that enables students to continually develop their skills in the interdisciplinary field of hybrid classical and quantum systems while being guided by experts in the respective fields. Our course approaches quantum-accelerated computing from an information systems perspective, which includes a new emphasis on different execution models and the design and analysis of hybrid quantum-classical applications on HPC systems.

Regarding future refinements of the course, the implementation of surveys could provide insights into the reception of our course and evaluate its effectiveness and areas for improvement with regard to defined learning outcomes. We also plan add components of peer instruction~\cite{MazurPeerInstruction} as a teaching approach to future iterations of the course to make the conventional lectures more interactive among students.

An exhaustive evaluation of challenges and opportunities in teaching QC to computer scientists to address future challenges and requirements of the post-Moore era, especially for HPC, could be a next step to extend the impact of our educational initiative.

Possible future developments of this course may also include the design of an advanced version of our \lecturename~course alongside an introductory-level course. While the introductory course may be leaning on already existing quantum computing introductions for computer scientists, the advanced course might build upon this foundation and delve into specific HPC-related concepts such as system design, software engineering, and the orchestration of HPC systems with regard to both traditional computing resources and quantum processing units. We also do not rule out reading tasks and an increased element of asynchronous learning in future iterations of the course.

Additionally, the long-term vision may include the establishment of an HPC-Quantum program at our university that consists of a number of courses dedicated to classical and quantum computing integration, in a direction similar to~\cite{thdegHighPerformanceComputing}.

\label{sec:conclusion}

\section*{Acknowledgement}

\label{sec:acknowledgement}

The development of the lecture is based on the knowledge gained through the following projects: "Runtime Control in Multi Clouds" (Rucon), Austrian Science Fund (FWF): Y904-N31 START-Programm 2015; Standalone Project "Transprecise Edge Computing" (Triton), Austrian Science Fund (FWF): P 36870-N; Standalone Project "Entanglement-Based Certification of Quantum Technologies" (EBCQT), Austrian Science Fund (FWF): P 36478-N funded by the European Union – NextGenerationEU; and by the Flagship Project "High-Performance Integrated Quantum Computing" (HPQC) \# 897481 Austrian Research Promotion Agency (FFG) funded by the European Union – NextGenerationEU. We would also like to thank the Austrian Federal Ministry for Digital and Economic Affairs, the National Foundation for Research, Technology and Development and the Christian Doppler Research Association. We would like to thank Dr. Olivia Vrabl for her valuable feedback on didactic aspects of the course.

We acknowledge the use of IBM Quantum services for this work. The views expressed are those of the authors and do not reflect the official policy or position of IBM or the IBM Quantum team.

\bibliographystyle{IEEEtran} 
\bibliography{references} 

\end{document}